\documentclass[sigconf]{acmart}

\setcopyright{none}
\renewcommand\footnotetextcopyrightpermission[1]{}

\AtBeginDocument{%
  \providecommand\BibTeX{{%
    \normalfont B\kern-0.5em{\scshape i\kern-0.25em b}\kern-0.8em\TeX}}}

\usepackage{booktabs,multirow}
\usepackage{siunitx}
\usepackage{tikz}
\usepackage{mathtools}

\newcommand{\anova}[4]{{\small $F_{#1,#2}=#3$, $p=#4$}}
\newcommand{\anovas}[3]{{\small $F_{#1,#2}=#3$, $p<.001$}}

\newcommand{\friedman}[3]{{\small$\chi^2(#1)=#2, p=#3$}}
\newcommand{\mannwithney}[2]{{\small$Z=#1, p=#2$}}

\usepackage{xspace}

\newcommand{\name}{DuSK\xspace}

\definecolor{author1}{rgb}      {0.9,0.5,0.0}
\definecolor{author2}{rgb}      {0.6,0.0,0.8}
\definecolor{author3}{rgb}      {0.0,0.4,0.5}
\definecolor{author4}{rgb}      {0.8,0.3,0.5}
\definecolor{author5}{rgb}      {0.9,0.0,0.0}
\definecolor{author6}{rgb}      {0.0,0.6,0.8}

\definecolor{yellowish}{rgb}  {0.5,0.4,0.0}

\newcommand{\mean}[1]{{\small M=#1}}
\newcommand{\std}[1]{{\small SD=#1}}
\newcommand{\mstd}[2]{{\small M=#1, SD=#2}}

\newcommand*\strokeone[2]{$
    \protect\vcenter{\protect\hbox{
        \protect\begin{tikzpicture}[scale=.22]
        \protect\draw[->, line width=0.25mm, align=center] (#1) -- (#2);
        \protect\end{tikzpicture}
    }}
$}

\newcommand*\stroketwo[3]{$
    \protect\vcenter{\protect\hbox{
        \protect\begin{tikzpicture}[scale=.22]
        \protect\draw[->, line width=0.25mm, align=center] (#1) -- (#2)  -- (#3);
        \protect\end{tikzpicture}
    }}
$}

\usepackage{csquotes}
\SetBlockThreshold{2}

\usepackage[normalem]{ulem}

\newcommand{\revDel}[1]{}

\usepackage[framemethod=tikz]{mdframed}
\usetikzlibrary{patterns}

\newlength{\hatchspread}
\newlength{\hatchthickness}
\newlength{\hatchshift}
\newcommand{\hatchcolor}{}
\tikzset{hatchspread/.code={\setlength{\hatchspread}{#1}},
         hatchthickness/.code={\setlength{\hatchthickness}{#1}},
         hatchshift/.code={\setlength{\hatchshift}{#1}},% must be >= 0
         hatchcolor/.code={\renewcommand{\hatchcolor}{#1}}}
\tikzset{hatchspread=3pt,
         hatchthickness=0.4pt,
         hatchshift=0pt,% must be >= 0
         hatchcolor=black}
\pgfdeclarepatternformonly[\hatchspread,\hatchthickness,\hatchshift,\hatchcolor]% variables
   {custom north east lines}% name
   {\pgfqpoint{\dimexpr-2\hatchthickness}{\dimexpr-2\hatchthickness}}% lower left corner
   {\pgfqpoint{\dimexpr\hatchspread+2\hatchthickness}{\dimexpr\hatchspread+2\hatchthickness}}% upper right corner
   {\pgfqpoint{\dimexpr\hatchspread}{\dimexpr\hatchspread}}% tile size
   {% shape description
    \pgfsetlinewidth{\hatchthickness}
    \pgfpathmoveto{\pgfqpoint{0pt}{\dimexpr\hatchspread+\hatchshift}}
    \pgfpathlineto{\pgfqpoint{\dimexpr\hatchspread+0.15pt+\hatchshift}{-0.15pt}}
    \ifdim \hatchshift > 0pt
      \pgfpathmoveto{\pgfqpoint{0pt}{\hatchshift}}
      \pgfpathlineto{\pgfqpoint{\dimexpr0.15pt+\hatchshift}{-0.15pt}}
    \fi
    \pgfsetstrokecolor{\hatchcolor}
    \pgfusepath{stroke}
   }
   
\tikzset{
} 

\mdfdefinestyle{strikeoutbox}{%
    linewidth=0pt,
  apptotikzsetting={\tikzset{mdfbackground/.append style={%
        pattern=custom north east lines,
        hatchspread=50pt,
        hatchcolor=red
        }
      }}
}

\begin{document}
% \input{tex/changes}

\title{DuSK: Faster Indirect Text Entry Supporting Out-Of-Vocabulary Words for Touchpads}


\renewcommand{\shortauthors}{}

\author{Damien Masson}
\email{damien.masson@umontreal.ca}
\orcid{0000-0002-9482-8639}
\affiliation{%
  \institution{University of Montréal}
  \city{Montreal}
  \state{Quebec}
  \country{Canada}
}

\author{Zhe Liu}
\email{zheliu92@cs.ubc.ca}
\orcid{0000-0002-1904-9045}
\affiliation{%
  \institution{University of British Columbia}
  \city{Vancouver}
  \state{British Columbia}
  \country{Canada}
}

\author{Charles Xu}
\email{qiang.xu1@huawei.com}
\orcid{0000-0002-0077-1003}
\affiliation{%
  \institution{Huawei HMI Lab}
  \city{Markham}
  \state{Ontario}
  \country{Canada}
}


\begin{abstract}
Given the ubiquity of SmartTVs and head-mounted-display-based virtual environments, recent research has explored techniques to support eyes-free text entry using touchscreen devices. However, proposed techniques, leveraging lexicons, limit the user's ability to enter out-of-vocabulary words. In this paper, we investigate how to enter text while relying on unambiguous input to support out-of-vocabulary words. Through an iterative design approach, and after a careful investigation of actions that can be accurately and rapidly performed eyes-free, we devise DuSK, a \textbf{Du}al-handed, \textbf{S}troke-based, \textbf{K}eyboarding technique. In a controlled experiment, we show initial speeds of 10 WPM steadily increasing to 13~WPM with training. DuSK outperforms the common cursor-based text entry technique widely deployed in commercial SmartTVs (8 WPM) and is comparable to other eyes-free lexicon-based techniques, but with the added benefit of supporting out-of-vocabulary word input.

% Typing or swiping to enter text on touchpads often results in noisy inputs given users’ inability to actively target on a display. A common solution to reduce the induced ambiguity is to rely on a lexicon against which to match words, however, this method does not permit out-of-vocabulary words. In this paper, we investigate how to enter text while relying on unambiguous input to support out-of-vocabulary words. Through an iterative design approach, and after a careful investigation of actions that can be accurately and rapidly performed eyes-free, we devise DuSK, a Dual-handed, Stroke-based, Keyboarding technique. In a controlled experiment, we show initial speeds of 10 WPM, reaching up to 30 WPM with training. DuSK outperforms the common cursor-based text entry technique widely deployed in commercial SmartTVs (8 WPM) and is comparable to other lexicon-based techniques, but with the added benefit of supporting out-of-vocabulary word input.
\end{abstract}



% Teaser figure
\begin{teaserfigure}
    \centering
    \includegraphics[width=0.9\textwidth]{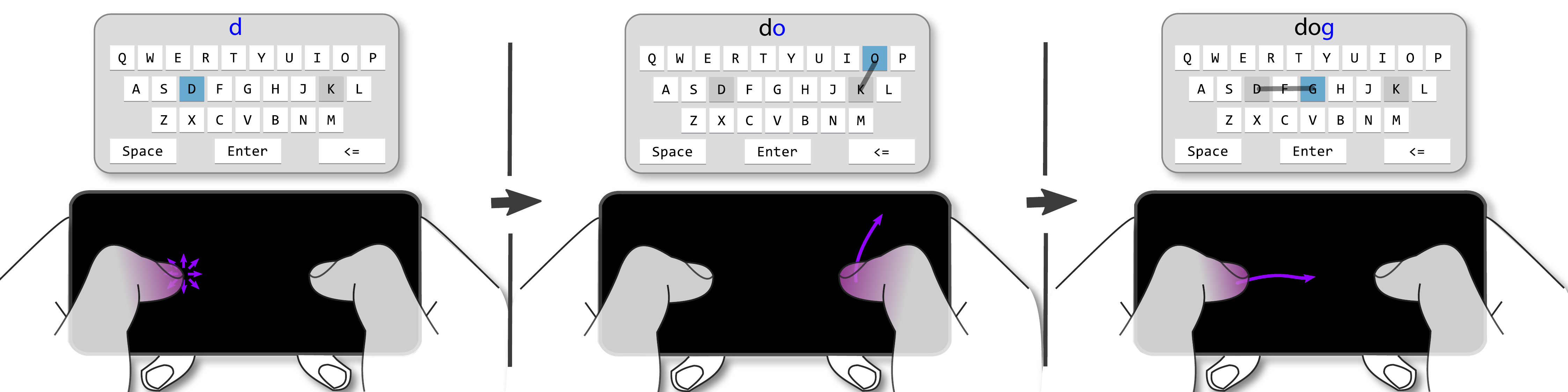}
    \caption{Entering `dog' with DuSK involves, with alternating thumbs, a tap ('D'), a short stroke (\strokeone{0, 0}{1, 1}`O'), a long stroke (\strokeone{0, 0}{2, 0}`G').
    \label{fig:teaser}}
    \Description{figure description}
\end{teaserfigure}

\maketitle

\section{Introduction}
Entering words without looking at the input device is desirable with SmartTVs or in Virtual Reality (VR)~\cite{blindtype, isfree, indirect_gesture_typing}. For text input on SmartTVs, decoupling the display from the input device eliminates the need to glance from display to input device, thus increasing focus and improving performance~\cite{peripheral_vision_typing, blindtype, isfree}.  It is also the only option when the input device or users' fingers are invisible, as in head-mounted-display (HMD) based VR ~\cite{vr_physical_keyboard, text_entry_selection_vr}.

One primary mechanism for supporting text entry in these contexts is through the use of touchpads, given that many remote controls are touch-enabled (e.g. Apple and Huawei SmartTV remotes, HTC Vive and Playstation controllers).
However, designing an eyes-free text entry method is challenging in the absence of physical keys; the lack of tactile and visual feedback combined with users' inability to monitor position aggravate precision and fat-finger errors~\cite{distracted_gestures_one_handed, PietroszekSpatial}. %Specifically, location-dependent gestures (ubiquitous when interacting with touchscreens) are difficult to perform without actively looking at the touchpad. 
For example, Lu et al. found that when users tap-type eyes-free, key regions formed by users' touch endpoints overlap considerably and typical text decoding algorithms are ineffective~\cite{blindtype}.

A popular solution to overcome noisy input is to infer the intended word using a probabilistic method. The algorithm searches a lexicon for words closest to users' input, optionally guiding its decision using previous inputs and the frequencies of words in natural text~\cite{gesture_typing, 10.1145/502716.502753}.
%Users then select the desired word among the most likely words yielded by the algorithm.
This approach boosts performance but can complicate the entry of out-of-vocabulary (OOV) words such as usernames, passwords, website addresses, proper names, and other desired character string components \cite{shrimp, velocitwatch, effects_language_modeling}. As such, two recent research efforts have explored leveraging smartphone-based text input techniques to support ``eyes-free'' input with external displays (e.g. SmartTVs or HMD-VR).  First, BlindType~\cite{blindtype} guesses likely characters given a user's tapping actions.  Users leverage their knowledge of character locations on a soft keyboard to estimate locations.  More recently, i'sFree~\cite{isfree} does exactly the same thing, but using word-gesture text entry instead of character-by-character tapping.  In both cases, a lexicon is used to identify the most likely word.  No support for OOV words is described.

Ideally, entering text -- even eyes-free -- would rely on unambiguous actions users can execute accurately, as on physical keyboards. While a lexicon could still support auto-correction and word completion \cite{antti_typing_in_the_wild} due to user imprecision, an ability to enter text deterministically would allow OOV words. The challenge then becomes how to design such a method while offering a fast input rate. As an example relevant to SmartTVs, the touchpad of AppleTV's remote controls a cursor over a virtual keyboard using unambiguous strokes and taps; this is essentially the same mechanism developed decades ago using physical arrow keys and an `OK' or `Enter' physical button on legacy remote controls.  And while this does allow the entry of OOV words, several studies showed that this technique and some variants reach a maximum input rate of 8~WPM~\cite{blindtype, remotes2, joysticks}. In comparison, sighted text entry on modern smartphones is up to four to five times faster~\cite{antti_typing_in_the_wild}.

In this paper, we investigate how to support efficient, eyes-free, touchscreen-based text entry that, via unambiguous actions, permits OOV words. 
As opposed to previous work in eyes-free text input~\cite{isfree, blindtype, escape_keyboard}, we explore the use of gestures performed with high accuracy coupled with statistical decoding approaches relying solely on users' input (i.e. no language model). This led us to propose and systematically evaluate a set of actions that can be executed quickly and accurately eyes-free. The results are helpful to inform the design of eyes-free techniques; as such, we proposed different designs based on these findings and informally tested them through an iterative process, culminating in the design of \name, standing for \textbf{Du}al-handed \textbf{S}troke-based \textbf{K}eyboarding technique. Our technique leverages two-thumb input~\cite{10.1145/1993060.1993066}, taps along the bezel~\cite{bezel_menu} and directional gestures~\cite{marking_menu} to support efficient text input, including OOV words, when the display is decoupled from the input device, such as in Virtual Reality or with SmartTVs.

\name can be viewed as two side-by-side ``regions'' where users can leverage short and long directional swipes to acquire individual characters in a deterministic fashion.  We present a summative evaluation of \name, demonstrating that new users can quickly achieve typing speeds of up to 13 WPM with deterministic input, while expert users reach speeds comparable to sighted tap typing on soft keyboards~\cite{antti_typing_in_the_wild}.

To summarise, our work makes the following contributions:
\begin{itemize}
    \item It reports on the results of an experiment proposing and evaluating eyes-free actions in order to inform the design of eyes-free techniques
    \item It presents the design and implementation of \name, a fast indirect text entry method that supports OOV words, and reports the insights collected along the way.
    \item It reports on the results of an evaluation of \name for both OOV and in-vocabulary words.
\end{itemize}

\section{Related Work}
For clarity's sake, performance metrics are excluded from the text and, instead, listed in Table~\ref{tab:summary}.

\subsection{Eyes-Free Text Entry}
In the literature, the term \textit{eyes-free} or \textit{sight-free} can refer to two different levels of feedback: 1) Users have some visual feedback such as the text entered on an external display (or head-mounted display) but cannot see their hands nor the input device~\cite{isfree, blindtype}; 2) Users rely solely on audio or tactile feedback~\cite{graffiti_eyes_free_error_correction}.
In line with other work in text entry~\cite{isfree, blindtype}, this work uses the former definition.

In this context, a significant body of work explored eyes-free input to external displays using input modalities such as touchpads~\cite{isfree, blindtype, indirect_gesture_typing}, TV remotes~\cite{remotes}, game controller joysticks~\cite{joysticks}, speech recognition~\cite{speech_tv}, accelerometers~\cite{gestext}, hand-tracking cameras~\cite{10.1145/2807442.2807504}, ray casting~\cite{text_entry_selection_vr},  smartwatches~\cite{10.1145/2909132.2909273}, sensors on the back of devices~\cite{sandwich_keyboard, back_of_device_typing}, or other handheld devices~\cite{rotoswype}.

Because of the ubiquity of the device, smartphone's touchscreen are often used as eyes-free input devices to external displays. Moreover, users' familiarity with smartphone-based text input can be leveraged to speed text input.  In this vein, Lu et al.~\cite{blindtype} proposed leveraging soft-keyboard tap typing. Because of the lack of precision from users when selecting small targets eyes-free~\cite{PietroszekSpatial}, they used statistical decoding coupled with a lexicon to disambiguate user input. Zhu et al.~\cite{isfree} adapted shapewriting~\cite{gesture_typing} for eyes-free usage. To compensate for variations in gesture locations, the imaginary-keyboard position is learned based on current and previous input.

However, while BlindType and i'sFree offer excellent performance (see Table~\ref{tab:summary}), they are limited to entering words present in their lexicon, as opposed to the classic text-entry method of moving a cursor over a virtual keyboard (cursor-based) using five keys (Up, Left, Right, Down and OK).  It is unclear how -- of even if -- these techniques could be adapted to support out-of-vocabulary (OOV) words.  As a result, while deterministic methods pale in comparison in terms of speed, because of the need to enter non-lexical words (passwords, websites, etc.), deterministic techniques remain the preferred method for text input on commercial SmartTVs (e.g. consider the Apple TV which, despite having a touchpad-based remote, uses touch-based five-key text entry as opposed to a more inferential, lexicon-based technique). 

\begin{table}[t]
\begin{tabular}{l|l|l|l|l|l}
\hline
\multirow{2}{*}{Technique} & \multirow{2}{*}{Method} & \multirow{2}{*}{MT} & \multirow{2}{*}{OOV} &  \multicolumn{2}{c}{WPM}\\

%\cmidrule(lr){5-6}
& & & & Start & End\\
\hline
i'sFree~\cite{isfree} & Gesture & No & No & 22 & 25 \\ \hline
BlindType~\cite{blindtype} & Tap  & No & No & 21 & 23 \\ \hline
Escape-Kb~\cite{escape_keyboard} & Menu & No & No & 7 & 15 \\ \hline
Bezel menus~\cite{bezel_menu} & Menu & Yes & Yes & 5 & 12 \\ \hline
Cursor-based~\cite{blindtype} & Swipe & No & Yes & 7 & 8 \\ \hline
Graffiti~\cite{graffiti_eyes_free_deterministic} & Gesture & No & Yes & 7 & 8 \\ \hline
\end{tabular}
\caption{\label{tab:summary}Summary of techniques for indirect text entry on touchpad. MT: Supports Multi-Touch. OOV: Out-of-vocabulary / do not rely on a lexicon. WPM: Words-per-minute for first (\textit{Start}) and last (\textit{End}) block.}

\end{table}

\subsection{Supporting Out-of-Vocabulary words}
When the input signal is ambiguous—as is the case with eyes-free input—text entry systems rely on a \textit{disambiguation strategy}. Commonly, noisy tap locations are clarified using probabilistic methods, such as Bayesian models~\cite{10.1145/502716.502753}, or machine-learning approaches that dynamically re-estimate key locations while typing~\cite{sandwich_keyboard, back_of_device_typing}. To further improve accuracy, some models incorporate a lexicon, reducing the need for users to take extra steps to clarify their intent~\cite{shrimp, blindtype}. However, without support for out-of-vocabulary (OOV) words, text entry remains limited to the words contained within this lexicon.%We see auto-correction across text entry systems, including both modern word processors and smart devices.

%Providing a secondary text entry mechanism can help support OOV words.
Wang et al. and Vertanen et al. showed that users can predict words that will be difficult for the decoder and decide on the strategy to use~\cite{shrimp, velocitwatch}. They then leverage a secondary text mechanism to support OOV words.  In SHRIMP~\cite{shrimp}, users have the option to tilt the phone to select a character deterministically instead of using linguistic disambiguation. Using the same idea, Vertanen et al. \cite{velocitwatch} found that a user could type on a watch more accurately, albeit slower, when anticipating a word to be difficult for the decoder. Finally, word-gesture keyboard users on smart-devices can switch to tap typing when faced with OOV words.

Unfortunately, state-of-the-art techniques to enter text on a remote display such as BlindType and i'sFree do not offer secondary mechanisms to support OOV words~\cite{blindtype, isfree, indirect_gesture_typing}. To be clear, the solution is not to ask users to interact more carefully as these techniques are fully dependent on lexicons; i'sFree is an adaptation of shape writing which does not allow OOV~\cite{gesture_typing} and BlindType leverages the thumb's muscle memory while typing common words -- as opposed to OOV words -- and requires users to select words within a list generated from a lexicon. Even if BlindType was modified to also let users enter out-of-lexicon words, this would require users to reach a tapping accuracy similar to sighted-typing. However, as Lu et al. demonstrated through their first study, users' tap-type locations significantly overlap in space in the absence of visual targeting~\cite{blindtype}. In fact, they report user inaccuracy as the motivating factor for using a  statistical decoding algorithm that includes a language model. Our work aims to allow the entry of all words, including OOV words, thus using statistical decoding methods that do not rely on a language model.

\subsection{Gesture-based Text Entry}
Myriad of gesture-based text entry techniques have been proposed on tactile surfaces. We classify them based on their encoding of each atomic action (e.g. a continuous stroke or through multiple discrete strokes~\cite{10.1145/2642918.2647354}), and in the level that they operate in: character-level, syllable-level or word-level.

Shape writing, also called \textit{SHARK2}, \textit{Gesture Keyboard} or \textit{Gesture Typing} is a popular, continuous, word-level text entry technique that often outperforms regular soft-keyboards~\cite{gesture_typing}. 
Indeed, operating at word-level often results in high speed at the cost of less expressivity: words not in the lexicon are harder to enter~\cite{antti_typing_in_the_wild}.

Full expressivity often means some form of character-level text entry. Handwriting is compelling considering that users are already familiar with it; however, the complexity of shaping and recognizing letters hampers its performance. For this reason, Goldeberg et al. proposed Unistroke~\cite{unistroke} and \textit{Palm,  Inc.} created Graffiti, which are both simplified alphabets whose usage result in faster character-level text entry than printing characters but require a learning curve.
Another trend of techniques, inspired by marking-menus~\cite{marking_menu}, investigates discrete character-level entry. Chen et al. proposed SwipeBoard~\cite{10.1145/2642918.2647354} for ultra-small devices; users enter a character using two consecutive swipes (or taps) first to select a region and then a character within that region. On smartphones, Banovic et al. presented EscapeKeyboard~\cite{escape_keyboard} for one-handed use, which leverages the Escape selection technique~\cite{escape}. These techniques do support OOV, but due to their unfamiliarity, they require some training from users.

%\subsection{Two-Handed Text Entry} % Maybe focus on handheld touchpads only
%While the most common and effective ``eyes-on'' text entry techniques use both hands, e.g. physical~\cite{how_we_type} and soft keyboards~\cite{antti_typing_in_the_wild, antti_kalq}, all the aforementioned variants were designed for one-finger usage, and the hand not interacting is either not used~\cite{blindtype, escape_keyboard}, or restricted to holding the input device~\cite{graffiti_eyes_free_deterministic, 10.1145/2642918.2647354}. Considering the benefits of involving both hands, some techniques have been proposed to support bi-manual eyes-free typing.
%Perkinput~\cite{perkinput} and BrailleTouch~\cite{brailletouch} are examples of such techniques, although they were designed for blind-users and rely on Braille and extensive training. 
%Shape writing~\cite{gesture_typing}, mentioned earlier, has been converted for two-thumbs usage and has shown encouraging results despite a slight drop in performance~\cite{bimanual_gesture_typing}. However, it was not designed for eyes-free interaction, and, like one-handed shape writing, cannot be used to enter out-of-lexicon words.
%Jain et al.~\cite{bezel_menu} proposed the use of bezel menus, leveraging the form factor of smartphones to enter text with two-hands but the technique again requires training and \textit{some} visual monitoring of the input device.
%Finally, Shi et al.~\cite{toast} proposed a method to touch type using 10 fingers, eyes-free, on a flat surface; however, their results do not apply to handheld touchpads.

\section{Design goals}
We design \name as a character-by-character eyes-free text entry method for touchpad-enabled controllers (e.g. the Apple TV remote, or a commodity smartphone used for input to a SmartTV or a HMD-based VR environment).

\name needs to work under a set of constraints imposed by the context (e.g. eyes-free, touchpad), to fill the gap left by related work regarding out-of-vocabulary words, and to respond to common requirements expected from text entry techniques (e.g. performance and learnability). Below, we summarize these five main design goals:

\begin{itemize}
    \item \textbf{Expressive:}
    Users need to enter words which are not always included in the dictionary (e.g. passwords, web addresses, proper names, etc.~\cite{shrimp}). While methods relying on lexicons provide excellent performance~\cite{antti_typing_in_the_wild}, unlike BlindType~\cite{blindtype} or i'sFree~\cite{isfree}, our technique must support out-of-vocabulary words and offer similar performance (in terms of WPM and error rate) on a corpus including OOV words as on a corpus restricted to in-vocabulary words.
    
    \item \textbf{Efficient:} A primary design goal of a text entry method is to let users enter text rapidly (high words-per-minute) and accurately (low error rate). As a baseline, cursor-based techniques, commonly used on SmartTVs and other commercial devices because they allow users to enter out-of-vocabulary words, have speeds of up to 8~WPM~\cite{blindtype} versus 23~WPM for dictionary-based techniques~\cite{blindtype, isfree}. Our goal is performance significantly better than baseline cursor-based techniques (8 WPM), preferably at speeds closer to BlindType~\cite{blindtype} and i'sFree~\cite{isfree} (23~WPM).
    
     \item \textbf{Eyes-free:}
    Numerous handheld devices only have a touchpad (e.g. remotes), as opposed to a touchscreen, and, in some scenarios, users cannot see the input device (Virtual Reality) or looking at it can be uncomfortable and reduce input speed (SmartTVs)~\cite{blindtype, isfree}. Our technique should support eyes-free usage, in which the user looks at an external display rather than at the input device and the technique should be usable (competitive WPM and accuracy) even if the device and users' hands are hidden.
    
    \item \textbf{Familiar:}
    Our technique should utilize familiar mechanisms to enable efficient knowledge transfer, allowing novice users to achieve competitive performance with minimal practice. Similar to previous work~\cite{blindtype, isfree, 10.1145/2642918.2647354}, this is best achieved by adopting well-known layouts (e.g., QWERTY keyboard).
    
    \item \textbf{Two-handed:} Users typically use both hands on physical keyboards (and sometimes on soft keyboards), as bi-manual interaction often leads to improved performance~\cite{antti_typing_in_the_wild} by enabling parallel finger movement. Although this can be challenging in handheld scenarios~\cite{10.1145/1993060.1993066}, our approach investigates bi-manual input to confirm that alternating thumbs enhances typing speed. We center our design on landscape-oriented touchpads and two-thumb interaction, a comfortable and effective two-handed posture~\cite{bezel_menu}.
\end{itemize}

Several challenges have to be overcame in order to support our design goals. First, to enable the entry of OOV words, our technique must allow letter-by-letter entry using unambiguous actions. Second, the action set must be large enough to offer a mapping with all required characters. Third, actions must be quick to support fast entry, but actions must also be accurately performed eyes-free.
We gather from previous work in eyes-free input~\cite{distracted_gestures_one_handed, distracted_input} and in spatial correspondence targeting~\cite{bezel_menu, PietroszekSpatial} a set of actions and evaluate their viability in our eyes-free, dual-handed context in order to design a text entry technique.

%Previous work in eyes-free input~\cite{distracted_gestures_one_handed, distracted_input} and in spatial correspondence targeting~\cite{PietroszekSpatial} has provided some guidance for users' ability to accurately swipe and tap on touchpads.  
%Additionally, bezels are almost always easy regions of the device to locate via touch alone when interacting~\cite{bezel_menu}.
%Given past work, our goal is to design a dual-handed, stroke-based, keyboarding technique.  We leverage a marking menu-like interface for individual character assignment~\cite{marking_menu, distracted_input}, but also support word completion, correction, and other discrete actions through spatial correspondence near the edges of the touch area~\cite{bezel_menu, PietroszekSpatial}.  With this in mind, we begin our design with a formative study that informs initial designs for stroking and tapping input.
\section{Study 1 - Eyes-Free Gesture Set}
The aim of this first study is to propose a list of actions that can be mapped to a character in order to support letter-by-letter text entry.
We then report how long it took (time) and how well (accuracy) participants performed these actions in an eyes-free task in order to inform the design of the text entry technique.

\subsection{Task 1: Taps and strokes}
We consider a large number of actions given that English text entry requires a set of at least 28 actions (26 letters in the alphabet, space and backspace). All these actions need to be unambiguous, fast, and reliably achieved eyes-free. Therefore, we draw from previous work in closely-related contexts such as distracted input~\cite{distracted_input, distracted_gestures_one_handed}, dual-handed marking menus~\cite{10.1145/1993060.1993066} and spatial targeting~\cite{PietroszekSpatial, bezel_menu}.
In particular, we include unistroke gestures as they are location-independent, thus easier to perform eyes-free. Following Bragdon et al.'s recommendation, we only consider mark-based unistroke gestures (originating from marking-menus, e.g.\strokeone{0,0}{1,1}, \stroketwo{0,1}{0.5,0.5}{0,0}) over free-from gestures as they are faster and more accurate~\cite{distracted_gestures_one_handed}. Further, to account for hand-preferences and differences amongst mark-based gestures due to thumbs' constraints~\cite{fat_thumb_boring, 10.1145/1993060.1993066}, we systematically include all \ang{45} and \ang{90} strokes as well as compound-strokes (two levels).
Finally, we also examine taps along the bezel; because of the device's form-factor, taps along the bezel are easier to perform eyes-free~\cite{distracted_gestures_one_handed, bezel_menu}. We follow Mohit et al.'s suggestion and divide the bezel into eight regions, essentially dividing the touchpad into 9 cells, with the central one unused. In total, we tested the following 64 actions:

\begin{itemize}
    \item \ang{45} and \ang{90} single-strokes (8 directions in total, e.g.\strokeone{0,0}{1,1},\strokeone{1,0}{0,0},\strokeone{1,0}{0,1}).
    \item L-Shape / \ang{90} compound-strokes (24 in total, e.g.\stroketwo{0,0}{1,0}{1,1},\stroketwo{0,1}{0.5,0.5}{0,0},\stroketwo{1,1}{1,0}{0,0}).
    \item V-Shape / \ang{45} compound-strokes (24 in total, e.g.\stroketwo{0,0}{1,0}{0,1},\stroketwo{0,0}{0,1}{1,0},\stroketwo{0,0}{1,1}{0,1})
    \item Taps along the bezel (8 in total, 3 top edge, 3 bottom edge, 1 left side, 1 right side, see Figure~\ref{fig:tap_locations}).
\end{itemize}
We further distinguish strokes by detecting which thumb is used based on each stroke’s starting location~\cite{10.1145/1993060.1993066}, effectively doubling the number of identifiable strokes and bringing the total to 120 unique actions. Additionally, we tested with two slightly different touchpad sizes to confirm that our results generalize across variations in device dimensions.

\subsubsection{Participants and Apparatus}
We recruited 14 participants (21 to 42 age range, mean = 27.14, 4 identified as female and 10 identified as male, 3 left handed). All but one were smartphones users, and only one participant used gesture typing frequently. We emulate a touchpad by using a smartphone that does not display any information.  All information was, instead, depicted on a 27-inch computer monitor positioned in front of the participants.
As smartphone, half the participants used an Honor Play (display size of 6.3 inches) and the other half a Huawei Mate 10 (5.9 inches). The experimental software was implemented in Java and communication between the smartphone and the computer connected to the display was done over UDP using the TUIO protocol\footnote{\url{https://www.tuio.org/}}.

\begin{figure}
    \centering
    \begin{minipage}{.49\textwidth}
        \centering
        \includegraphics[width=1\linewidth]{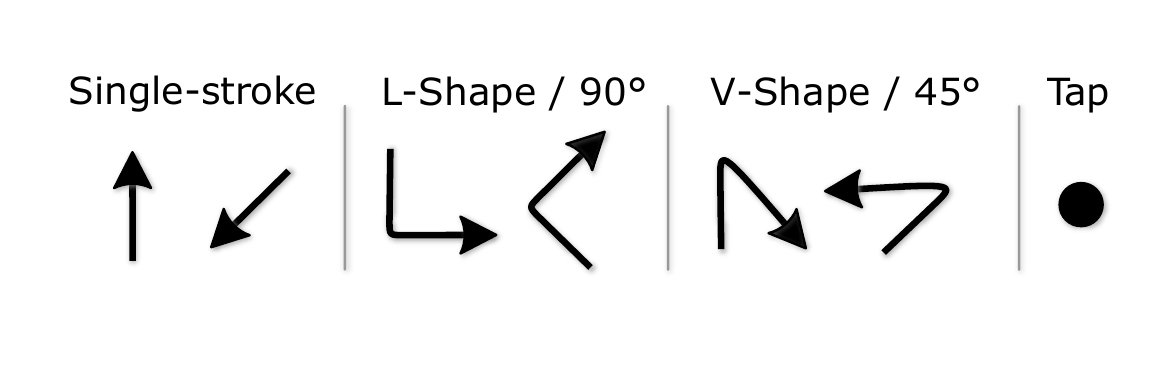}
        \captionof{figure}{Example of visual stimuli during Task 1}
        \label{fig:stimuli_strokes}
    \end{minipage}
    \hfill
    \begin{minipage}{.49\textwidth}
        \centering
        \includegraphics[width=1\linewidth]{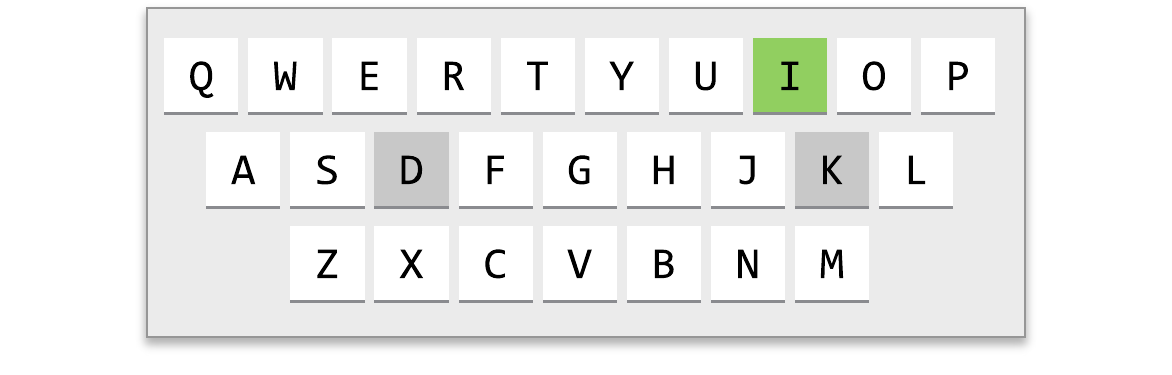}
        \captionof{figure}{Example of visual stimuli during Task 2, here, the participant is asked to stroke towards `I'}
        \label{fig:stimuli_keys}
    \end{minipage}
\end{figure}

\subsubsection{Design and Procedure}
We used a 2x2x4 mixed-design with the following factors and levels: {\sc Touchpad size} (between-subject, {\sc 6.3inches} or {\sc 5.9inches}), {\sc Thumb} (within-subject, {\sc Left} or {\sc Right}) and {\sc Action} (within-subject, {\sc Single-Stroke}, {\sc L-Shape Stroke}, {\sc V-Shape Stroke} or {\sc Taps}).

Participants sat in front of the computer display. They were asked to hold the smartphone horizontally (in landscape mode) under the desk so that they could not see nor visually monitor the phone, and to use both thumbs to perform strokes; an experimental design technique adapted from~\cite{distracted_input}.  The goal is to eliminate the confound of peripheral visual monitoring of position of the handheld device.

An action was shown on the display (see Figure~\ref{fig:stimuli_strokes}) either on the left or the right of the display. Participants were asked to reproduce the action using the appropriate thumb (e.g. left thumb if the gesture is shown on the left). When an action was done with the wrong thumb (we detect the thumb used based on the location of the first touch, left side means left thumb), the stroke was not registered and the application prompted participants to try again. When the action was completed, the next trial was immediately displayed.

%\textit{Task 2.} A keyboard layout identical to the default iPhone keyboard was displayed at all times on the display in front of the user (Figure~\ref{fig:stimuli}.2). Participants were instructed to imagine that their strokes started on `D' when using their left thumb and `K' for the right thumb (both keys highlighted in gray). For each trial, the key that had to be selected was highlighted in green. No feedback was given on the performed gesture and participants were free to use either thumb.

Participants performed actions in a random order. In the end, we obtained the coordinates of both thumbs when in-contact for ((56 {\sc Strokes} x 2 {\sc Thumbs} + 8 {\sc Taps}) x 2 {\sc Repetitions}) x 14 {\sc Participants} = 3360 {\sc Actions}.

\subsubsection{Measurements} We measure \textit{Time} as the time in milliseconds between the "DOWN" event that started a gesture and the corresponding "UP" event, as received by the smartphone. \textit{Accuracy} is measured using a recognition algorithm working as follow: $G_{ref}(i)$ corresponds to the i-th coordinate of the reference gesture (i.e. the action that was shown on the display for the participant to reproduce), and $G_{user}(j)$ corresponds to the j-th coordinate composing the participant's gesture. We first distinguish taps from strokes by measuring the sum of the distances of all the pair of coordinates composing $G_{user}$. A distance of less than 10mm (found through trial and error) is a tap, and everything higher is a stroke. Then, we use two different algorithms:
\begin{itemize}
    \item For taps: the touchpad is divided in 9 equally sized cells. The tap is recognized if $G_{user}(0)$ lies within the cell of $G_{ref}(0)$~\cite{PietroszekSpatial}.
    \item For strokes: we compute the ``deviation'' of the participant's gesture from all 56 tested strokes. The deviation is measured using the Dynamic Time Warping distance between the re-sampled (n=10) sequence of angles of $G_{user}$ and the angles of the stroke tested against. The stroke is accurately recognized if the deviation between $G_{user}$ and $G_{ref}$ is the lowest of all computed deviations.
\end{itemize}

\subsection{Task 1: Results and Discussion}
We used a repeated measure ANOVA with Greenhouse-Geisser correction when Sphericity was violated. The normality assumption of the data was verified using Q-Q plots.

\textbf{Time.} We found a significant main effect for {\sc Action} (\anovas{3}{36}{53.99}). Examining averages, we found that taps are the fastest (\mstd{151ms}{96}), followed by single directional strokes (\mstd{321ms}{234}). L-shape strokes (\mstd{708ms}{366}) and V-shape strokes were the slowest (\mstd{738ms}{396}).

\begin{figure}
    \centering
    \begin{minipage}{.49\textwidth}
        \centering
        \includegraphics[width=1\linewidth]{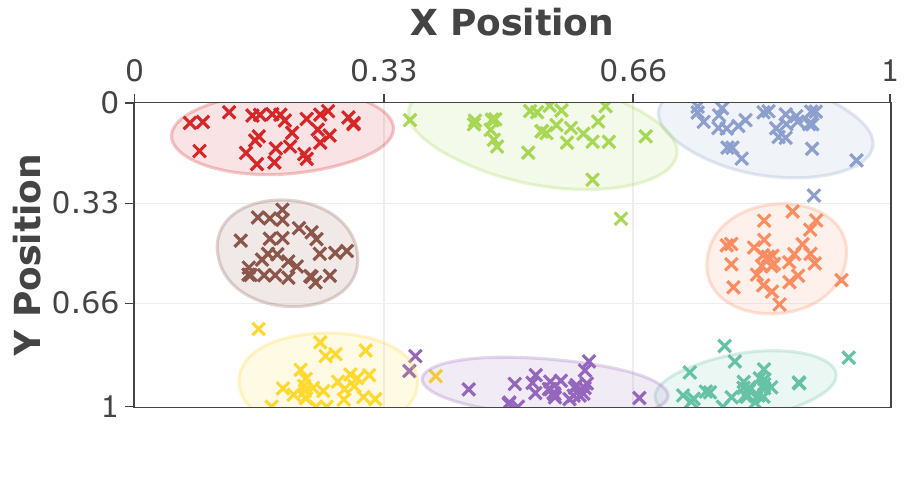}
        \captionof{figure}{Normalized locations of taps and 95\% confidence ellipses}
        \label{fig:tap_locations}
    \end{minipage}
    \hfill
    \begin{minipage}{.49\textwidth}
        \centering
        \includegraphics[width=.49\linewidth]{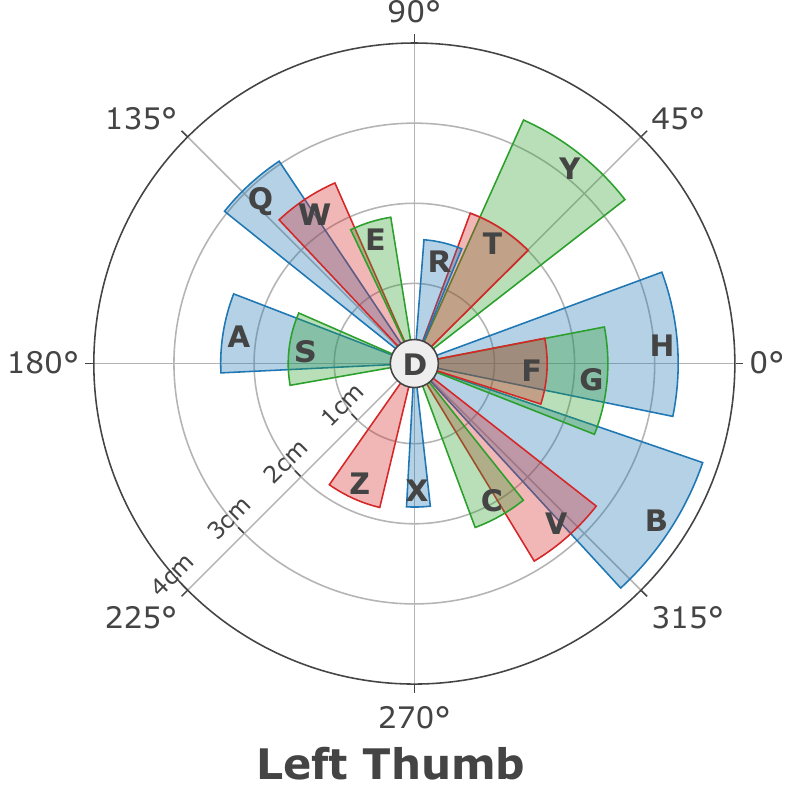}
        \includegraphics[width=.49\linewidth]{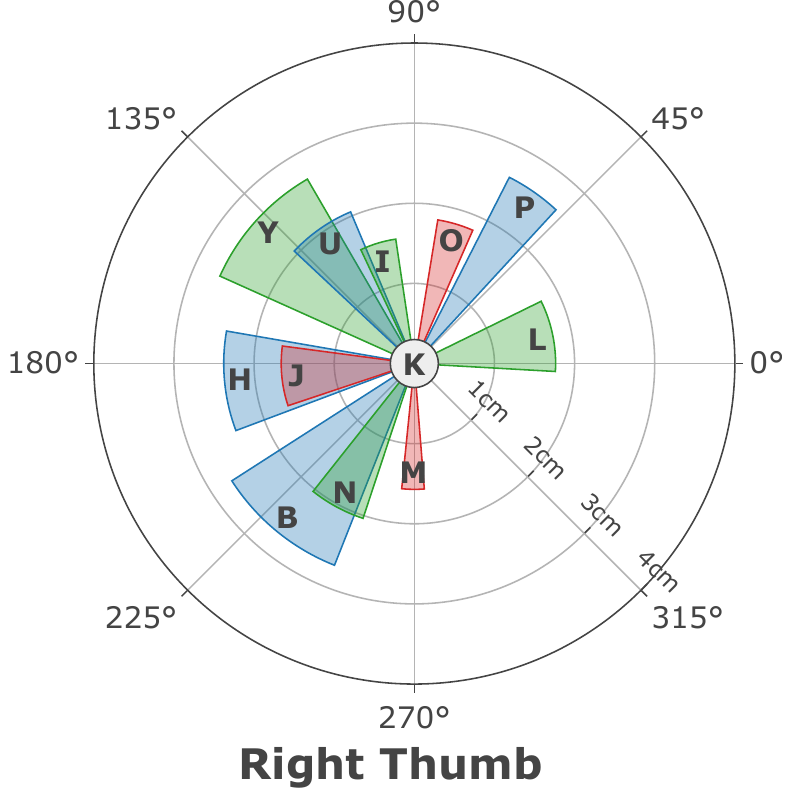}\hfill
        \captionof{figure}{Mean angle and length of strokes for each key, range corresponds to standard deviation x 2}
        \label{fig:mean-angle}
    \end{minipage}
\end{figure}

\textbf{Accuracy.} We found a significant main effect for {\sc Action} (\anovas{3}{36}{68.34}). Taps were the most accurate (\mstd{98\%}{13}, see Figure~\ref{fig:tap_locations}), V-Shape strokes were the second most accurate (\mstd{77\%}{42}), followed by L-Shape strokes (\mstd{71\%}{46}). Finally, single strokes were the actions performed with the lowest accuracy (\mstd{63\%}{48}). Interestingly, with the exception of taps, this order is reversed compared to \textit{Time}, suggesting that there is a trade-off speed/accuracy to consider. We also observed large differences among straight single-stroke (accuracy in descending order, high is better: \strokeone{0, 1}{0, 0}~100\%, \strokeone{0, 0}{0, 1}~98\%, \strokeone{0, 1}{1, 0}~70\%,
\strokeone{1, 0}{0, 1}~63\%,
\strokeone{0, 0}{1, 0}~48\%, 
\strokeone{1, 1}{0, 0}~45\%,
\strokeone{1, 0}{0, 0}~41\%,  \strokeone{0, 0}{1, 1}~38\%).

\textbf{Effect of thumb.} We did not find a significant effect of {\sc Thumb} on \textit{Time} (\anova{1}{12}{2.63}{.131}), however, we found a significant effect on \textit{Accuracy} (\anovas{1}{12}{14.19}). Actions done with the right hand were, on average, performed more accurately (\mstd{78\%}{41}) than actions done with the left hand (\mstd{72\%}{45}).
%TODO: Add comparison on accuracy based on the "quarters"

\textbf{Effect of touchpad size.} We did not find a significant of {\sc Touchpad Size} on \textit{Time} (\anova{1}{12}{2.85}{.117}) nor \textit{Accuracy} (\anova{1}{12}{0.58}{.461}). This suggests that tested actions are robust to slight variations of touchpad sizes.

In summary, our results suggest that gestures done on a smartphone held horizontally are accurate even if they are performed eyes-free. Straight directional strokes are performed faster but not necessarily more accurately than compound strokes. Additionally, strokes with a right-angle should be preferred over \ang{45} angles. Finally, slight variations in size of the input device does not seem to impact the time and accuracy of the gestures.

\subsection{Task 2: Mapping keys to strokes}
Through a second task, we look at how participants naturally associate strokes to keys. The objective is twofold: first, it informs us on how users ``map'' strokes to keys, and which thumb they associate to each key, given that they can use both thumbs. Second, it gives us a better idea of the level of precision that can be achieved when participants aim for a specific key eyes-free and using directional strokes.
This second task used the same participants and apparatus as the first task. Differences in the procedure are reported below.

\subsubsection{Procedure}
Similar to Task 1, participants were seated in front of the display and instructed to hold the smartphone horizontally and out of view under the desk. A keyboard layout matching the default iPhone keyboard was continuously displayed on the screen in front of them (Figure~\ref{fig:stimuli_keys}). Participants were asked to imagine that gestures made with their left thumb originated from the ‘D’ key and those with their right thumb from the ‘K’ key, which were highlighted in gray as a reminder (see Figure~\ref{fig:stimuli_keys}). By assigning fixed starting points for gestures, we aimed to limit the possible outcomes and obtain more meaningful results. We selected ‘D’ and ‘K’ because they are centrally located on the left and right sides of the keyboard, minimizing the distance required to reach other keys. During each trial, the target key to be selected was highlighted in green (see Figure~\ref{fig:stimuli_keys}). No feedback was provided on the gestures performed, and participants were free to use either thumb.

Keys were randomly ordered for each participant and repeated 10 times. In the end, we obtained the coordinates of both thumbs when in-contact for (24 {\sc Keys} x 10 {\sc Repetitions}) x 14 {\sc Participants} = 3360 {\sc Strokes}.

\subsection{Task 2: Results and Discussion}
Participants were not always consistent in their choice of thumb; some keys were selected using both the left and right thumbs. Participants also occasionally reported mistakenly selecting the wrong key. To address these errors, we excluded strokes for specific keys that were performed fewer than three times with a particular thumb. This adjustment resulted in a total of 3,313 valid trials, representing a retention rate of 98.6\% from the original 3,360 trials.

The average angle and length of each stroke towards a key aggregated over all participants can be seen on Figure~\ref{fig:mean-angle}. We observe some overlap suggesting that participants are not accurate enough to stroke towards certain keys reliably. However, we noticed that participants consistently varied the length of their strokes, depending on how far the key was.
As a result, when considering both lengths and angles of strokes, the selected key appears identifiable. To confirm this hypothesis, we implemented a recognition algorithm using the mean position of the normalized end point of each stroke of a key. A stroke is associated to a key based on the distance of its end-position to the mean end-position for each key in the training set. Using 10-fold cross validation, we obtained a mean top-1 accuracy of 68\%, and a top-2 accuracy of 93\%.

On average, it takes 346ms to select a key. Unsurprisingly, the farther away the key is from `D' or `K', the longer it takes to perform a stroke. Therefore, `L', `S', `J', `F', `X', `M' were the fastest keys to select, with times under 300ms, while `Y', `Q', and `B' were the slowest, with times exceeding 400ms. 
%To handle those missing cells, we used a Linear Mixed Effect Model with {\sc Participant} as random effect and {\sc Thumb} and {\sc Key} as fixed effect.

% Movement time
% Distance
% Effect of key/thumb on movement time and distance => Maybe some keys are too difficult to reach
\section{\name Design}
Informed by the preceding study, we design \name by following an iterative design approach;  we piloted each iteration, identified problems, and modified designs.  In this section, we provide an overview of these iterations.
%\subsection{Design Principles}
%The motivation behind \name is to propose a two-handed alternative to cursor-based text entry for handheld touchpads. As such, \name needs to offer the same capabilities as the cursor-based method while being easy to learn and resulting in better performances for both novice and expert users.
%Consequently, we designed the technique by following an iterative design approach focused on 4 design principles:

%\textit{Efficiency:} The technique should be fast, evaluated by Words-Per-Minute, and accurate, evaluated by error rate.

%\textit{Familiarity:} The technique should leverage familiar mechanisms to facilitate the transfer of knowledge and expertise.

%\textit{Expressivity:} The technique should allow for any type of word to be entered, including out-of-dictionary words, passwords, proper names, etc.

%\textit{Eyes-free:} Typing should be possible without looking at the input device; visual feedback only appears on the display.

% See gesture bar or Escape Keyboard
% For the principles "Examination of Text-Entry Methods for Tabletop Displays"

\subsection{Prototype Evolution} % Maybe not
%TODO: Say that we explore the 2 possibilities presented by Kin et al. (possibilities vs speed)
\begin{figure}[b]
    \centering
    \begin{minipage}{.49\textwidth}
        \centering
        \includegraphics[width=1\linewidth]{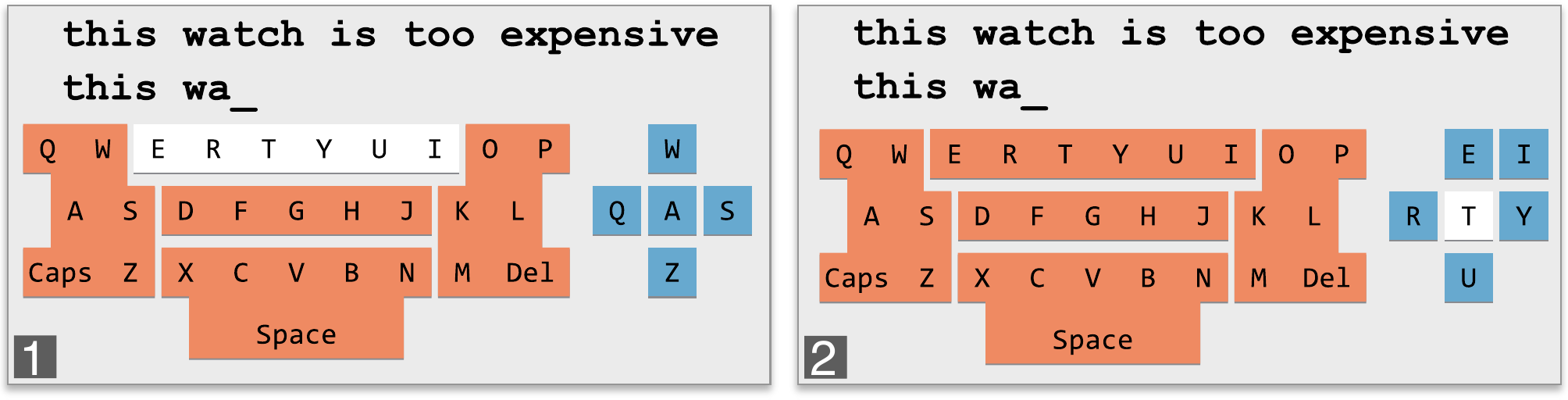}
        \captionof{figure}{Second design iteration, letters are selected in two steps e.g. To select `T': \strokeone{0, 0}{0, 1} with left thumb then tap with right thumb.}
        \label{fig:iteration_markingmenu}
    \end{minipage}
    \hfill
    \begin{minipage}{.49\textwidth}
        \centering
        \includegraphics[width=1\linewidth]{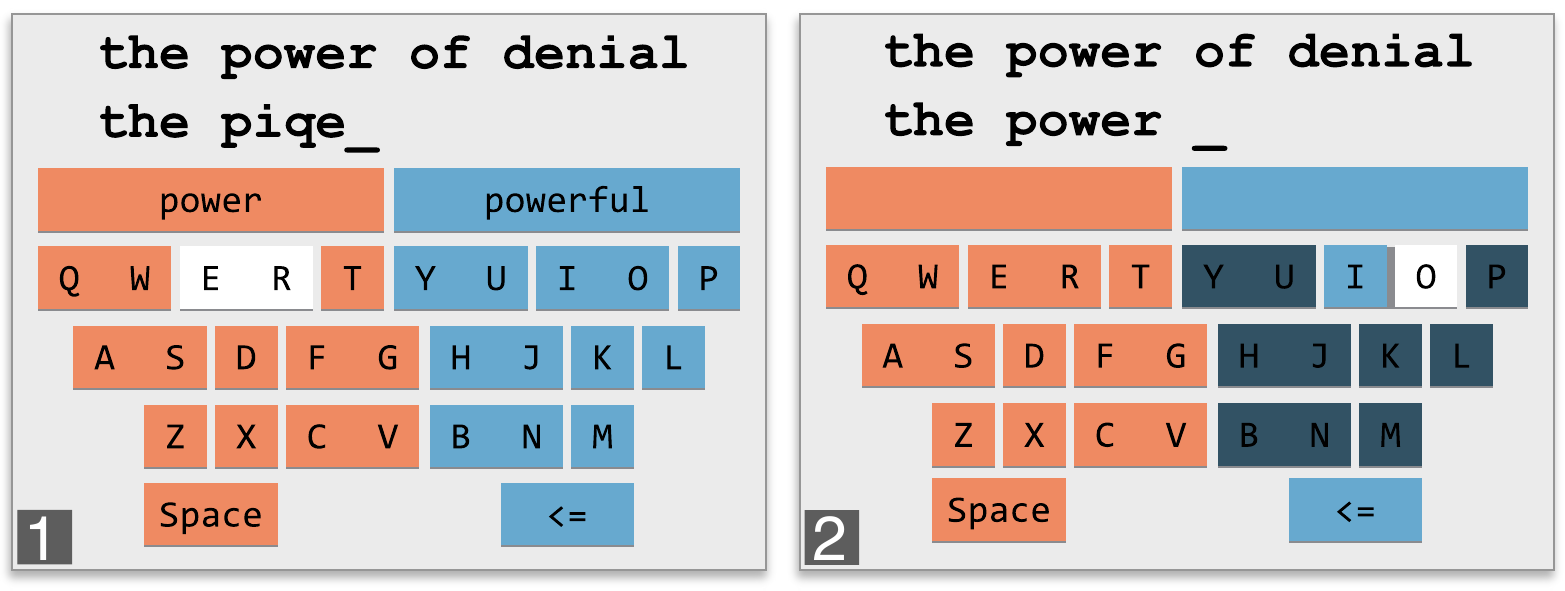}
        \captionof{figure}{Third design iteration, 1. To select `power' stroke towards [P] > [IO] > [QW] > [ER] > [ER]. 2. To select `O': \stroketwo{0, 0}{0, 1}{1, 1} with right thumb.}
        \label{fig:iteration_zonemenu}
    \end{minipage}
\end{figure}

\textbf{First iteration.} The prototype consisted of two 8-directional marking-menus~\cite{marking_menu}, each controlled by a thumb. A character was selected in two steps: first, by selecting a group of keys with the left thumb (with three to five characters per group), then by selecting a key within this group with the right thumb. Selections were done through single directional strokes, or a tap for the central item (similar to SwipeBoard~\cite{10.1145/2642918.2647354}), resulting in $9*9=81$ accessible characters.
We also added a predictive mode triggered when the second step is skipped (i.e. by only selecting group of keys). In that case, the right thumb selects suggestions generated by an algorithm similar to \textit{T9\footnote{\url{https://en.wikipedia.org/wiki/T9_(predictive_text)}}}. This allowed for faster entry of dictionary-words, while conserving the ability to enter words letter by letter using the regular method.
A first informal test revealed that participants were too unfamiliar with the layout (arranged in a ring, roughly following a QWERTY layout) causing slow reaction times.

\textbf{Second iteration.} We correct users' initial confusion by arranging keys like a soft keyboard to increase familiarity. Moreover, we limit actions to only 4-directions in an attempt to boost performance, see Figure~\ref{fig:iteration_markingmenu}. A second informal test with participants revealed that they could quickly locate keys on the display but were still slow in determining the stroke to reach the key. Moreover, because the two marking-menus were used sequentially, the benefits of using two fingers for faster selection were largely underexploited~\cite{10.1145/1993060.1993066}.

\textbf{Third iteration.} In order to improve speed, we reduced the number of steps to select a letter to only one by assigning different sets of letters to each hand. While this allows for simultaneous and faster selections~\cite{10.1145/1993060.1993066}, it results in a lower number of accessible items: only 18 (9 per thumb). Consequently, we formed groups to fit all 26 letters into less than 18 items by manually optimizing four constraints: 
1) Letters are arranged like a QWERTY keyboard because of its familiarity; 
2) Letters are assigned to the hand recommended by touch-typing guidelines; 
3) The central item is a always a single letter relatively close to all other groups; 
4) Difficult strokes identified in Study 1 are assigned to a single letter (or none, if possible). 
This resulted in the layout shown in Figure~\ref{fig:iteration_zonemenu}.
We further increased the input vocabulary by leveraging the tapping areas identified during Study 1 to select suggestions, 'Space', and 'Backspace'.
Similar to the previous iteration, a word is predicted by consecutively selecting groups of letters (Figure~\ref{fig:iteration_zonemenu}.1). Additionally, L-shaped strokes (e.g.\stroketwo{0, 0}{0, 1}{1, 1}) select letters deterministically (Figure~\ref{fig:iteration_zonemenu}.2). This last method allows for the entry of out-of-vocabulary words.

Results from a pilot study with 3 participants showed encouraging results for the predictive method, about 15~WPM after a few minutes of training. However, participants reported confusing feedback and were sometimes lost while typing long words (we show the first letter of the selected group and correct it later, e.g. ``power'' shows as ``piqee'' before applying dictionary-disambiguation, see Figure~\ref{fig:iteration_zonemenu}.1). Additionally, a participant repeatedly tried to select a letter using the thumb that could not reach the key (e.g. the participant wanted to select 'G' using the right thumb). This confirms our finding from Study 1 that participants have preferences regarding which thumb to use, similar to hand preferences observed from non-touch-typists on physical keyboards~\cite{how_we_type}. Finally, participants judged the deterministic method using L-shaped strokes to be \textit{unnatural} and entry rate reached around 8~WPM. While we were expecting these gestures to be slower (see Study 1), we did not anticipate participants having to pause and think mid-way through strokes. Participants would most likely improve through practice and muscle memory, but this goes against our goal of minimizing the learning curve.

\begin{figure}
	\centering
	\includegraphics[width=.477\textwidth]{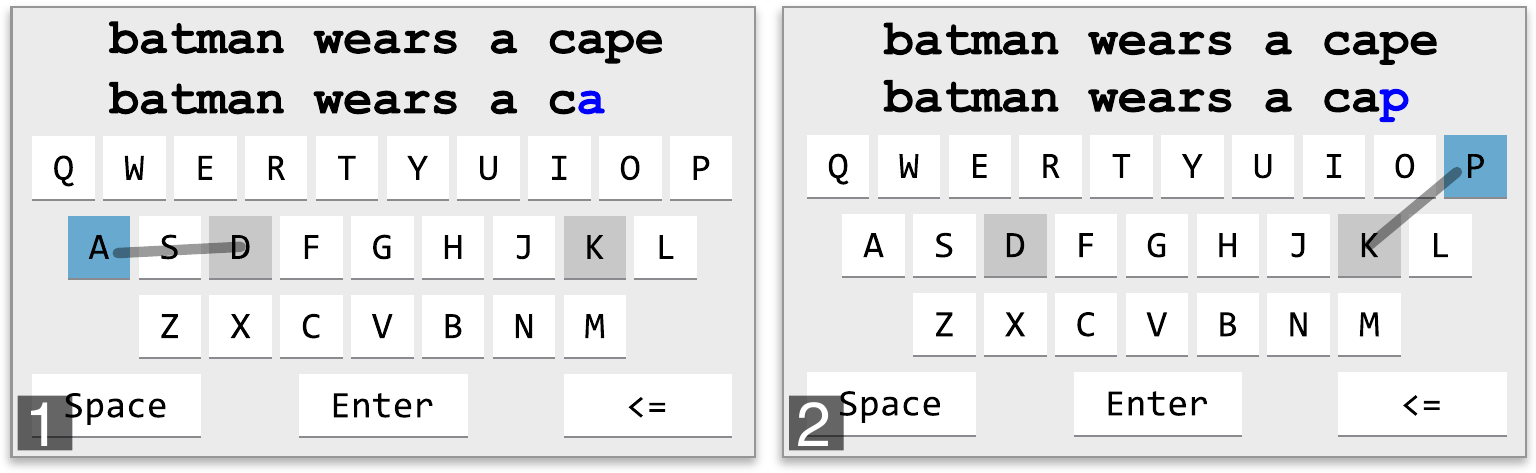}
	\caption{Final design iteration, 1. To select `A': long stroke with left thumb (\strokeone{2, 0}{0, 0}). 2. To select `P': stroke (\strokeone{0, 0}{1, 1}) with right thumb.}
	\label{fig:iteration_dualstrokekeyboad}
\end{figure}

\subsection{Insights from Iterative Design}
From the iterative design approach, we draw the following key insights in order to design \name's final iteration:
\begin{enumerate}
    \item \textbf{Familiar layout:} We reached a similar conclusion to Banovic et al.~\cite{escape_keyboard} in that using a familiar layout such as QWERTY helped participants. We see the choice of layout as a trade-off on typing speed between high lower-bound and high upper-bound; an optimized layout might result in higher expert-performance, while a familiar one will improve initial performance. We believe that a familiar layout is preferable for \name as our prototypes showed that using strokes to select characters already induces a substantial amount of practice by itself which we do not want to aggravate.
    \item \textbf{Same input dynamics:} Prototypes using a different mechanism to enter OOV words caused confusion for participants~\cite{williamson2006continuous}.  Participants had difficulties using both mechanisms and did not seem to transfer their experience from one to the other, even when both mechanisms were similar.
    \item \textbf{Respect hand-preferences:} Preferences regarding the hand to use to access a character are commonly observed with bi-manual text-entry techniques~\cite{how_we_type, how_we_type_mobile}. We found that participants were frustrated to be forced to use and remember a specific hand to select a character.
    \item \textbf{Breadth over depth:} While previous research on marking-menus found breadth and depth to be an even trade-off~\cite{marking_menu}, we found our participants to prefer breadth. Compound-strokes (depth of two) were slower than anticipated as they were often performed in two times (with a pause for visual search), while single strokes, even with a higher breadth (e.g. more angles) were faster and did not appear to cause a higher error rate.
\end{enumerate}

%\revAdd{
%We concluded that, in accordance with previous work~\cite{escape_keyboard, bezel_menu}, users are not accustomed to using strokes as a way of selecting characters -- as opposed to tap-typing or shape writing -- resulting in low initial performance. As a trade-off, past the first iteration, we preferred using a familiar keyboard layout (QWERTY) instead of an optimized one, resulting in improved initial performances at the cost of potentially lower expert performances. Second, all our attempts at providing a second text entry mechanism to enter OOV words confused participants. Moreover, participants had difficulties using both mechanisms and did not seem to transfer their experience from one to the other. Third, participants were frustrated to be forced to use and remember a specific hand to select a character. Finally, we found compound-strokes to be slower than anticipated as they are often performed in two times (with a pause for visual search). We draw from these insights to design the final iteration of \name.
%}

\subsection{Final Design}
In \name's final design, we removed L-shaped strokes that were difficult for novices and relied solely on single directional strokes. We disambiguate the character selected by using strokes' length, following Study 1's finding that that participants had a natural tendency to vary the length of their strokes depending on the position of the key.
Therefore, only one character is associated to each key, and both the length and the angle of a stroke can be varied to reach a specific key (see Figure~\ref{fig:iteration_dualstrokekeyboad}). Also, to support users' hand preferences, we do not constrain them on which thumb to use -- they simply need to stroke ``farther'' to access more distant letters. Additionally, we added a visual feedback to indicate the thumb's current position on the keyboard.

Unlike early iterations which proposed two modes with different activation mechanisms, this iteration only supports a deterministic mode that can be augmented with word completion and correction. This solution has the advantage of supporting rehearsal as defined by Kurtenbach et al.~\cite{rehearsal_design_principles}: Given that the action required from novice users is identical to expert users, users can develop their expertise while using the technique. We expect novice users to carefully select each character, while expert users perform faster, relying on auto-correct to compensate for the decreased accuracy.  %As on soft keyboards, experts will also become familiar both with when they need to type with care (out-of-vocabulary words) and, as on state-of-the-art soft keyboards, we can include the deterministic word among the auto-correct alternatives.

In the rest of the section, we detail the specifics of \name's implementation. The source code of our implementation is available on GitHub at <supressed> \footnote{To preserve anonymity, posting on GitHub will occur after acceptance of this paper}.

\subsection{Strokes' starting position}
With \name, users vary the length of their stroke to reach keys. Therefore, by carefully selecting strokes' starting position, we can reduce the average distance to travel to reach keys. 

To choose strokes' starting position, we simulated all possible starting positions on either side and computed their average distance to other keys. On the left side, we choose `D' as the starting position as it is the closest key to all other keys on that side (\mean{1.58 key radius}). On the right side, both `J' and `K' have a similar distance to other keys (respectively \mean{1.49} and \mean{1.56}). We decided to use ‘K’ as it reduces the number of keys in the bottom right corner, a direction that was shown to be difficult to stroke with the right thumb~\cite{10.1145/1993060.1993066}.

\subsection{Transfer function}
Essentially, \name uses two cursors (one per thumb) which return to their respective positions (either `D' or `K') after each stroke.
Below, we detail how we obtained the transfer function~\cite{transfer_function} controlling these cursors. We define the transfer function as a function mapping touchpad coordinates ($(X_{t}, Y_{t})$) to coordinates on the display $(X_{d}, Y_{d})$:
\begin{equation}
f(X_{t}, Y_{t}) = (X_{d}, Y_{d})
\end{equation}
We compute the transfer function using the strokes collected during Study 1. For each stroke, we have its normalized ending position (i.e. $(X_s, Y_s) = endingPosition-startingPosition$) on the touchpad and the corresponding position on the display of the key that the participant was aiming for (noted $(X_{k}, Y_{k})$). To minimize the impact of outliers, we only keep strokes whose normalized ending position is less than two times standard deviation away from the average position collected for the corresponding key (92.1\% of all collected strokes).
We then model the relationship between the touchpad and display coordinates by applying a linear regression as follows.
\begin{equation}
    \begin{cases}
      X_k = X_s * a_x + Y_s * b_x + c_x\\
      Y_k = X_s * a_y + Y_s * b_y + c_y
    \end{cases}       
\end{equation}
We repeat the operation for both thumbs, resulting in a different transfer function depending on the thumb used (the thumb used is inferred based on the starting location of the stroke).
Finally, we use the coefficients obtained from the linear regression to compute  $(X_{d}, Y_{d})$ from $(X_{t}, Y_{t})$.

\subsection{Word correction and completion}
While auto-correct and word completion are not necessary to achieve good performance with \name, they have some benefits for common words (see Study 2) and are common with modern soft keyboards.
We describe in this section how such algorithms can be implemented while preserving \name's ability to enter OOV words.

Both our implementations for auto-correct and word completion rely on a model similar to Goodman et al.'s Bayesian model~\cite{venolia2001language}. Therefore, our model combines the word probability with the input probability: Given an input $I$, the probability that it corresponds to a word $W$ is defined as follows:
\begin{equation}
P(W|I) = P(I|W) * P(W)
\end{equation}

\subsubsection{Input probability}
The input probability is defined by the probability of a sequence of strokes. A stroke is defined by its normalized ending position (i.e. $endingPosition-startingPosition$). From Study 1, we compute the average stroke ending-position for each key and their covariance matrices. We then use a bivariate Gaussian distribution to model the probability distributions of an observed stroke. This gives us $P(W_i|I_i)$ where $W_i$ is the $i$th character of $W$, and $I_i$ the $i$th stroke of the input.
The input probability is defined as the product of the individual probabilities of the strokes.
\begin{equation}
P(I|W) = \prod_{i=1}^{n} P(W_i|I_i)
\end{equation}
Where $n$ corresponds to the number of characters in $W$.

\subsubsection{Word probability}
The probability of a word $P(W)$ is defined by its frequency count in the English language normalized by the sum of the frequency counts of all possible words. We compute the list of possible words by retrieving the 3 characters yielding the highest probability for each stroke forming the input. We then compute all the possible letter combinations. For example, an input formed by five strokes would result in $3^5=243$ letter combinations. Combinations that are not within the top 50,000 words from the frequency count dictionary of the \textit{American National Corpus}~\cite{ANC} the dictionary; what remains forms the list of possible words.

\subsubsection{Auto-correct}
After entering `Space', the word entered is replaced by the word with the highest probability. The word is not auto-corrected if none of the generated possible words was found in the dictionary (out-of-vocabulary). Pressing backspace just after auto-correct reverts the word to its original spelling.  %It is this feature that permits the typing of out-of-dictionary words, a facility not easily supported by BlindType \cite{blindtype} nor i'sFree \cite{isfree}.

\subsubsection{Word completion}
To generate suggestions, words in the dictionary that are prefixed by any of the possible letter combinations are added to the list and their probability computed. We use a Trie~\footnote{\url{https://en.wikipedia.org/wiki/Trie}} structure to do the search efficently. In our current implementation, we only show the two words with highest probability (top-2).
\section{Study 2 - Evaluating \name}
To estimate the performance of \name, we evaluate the technique in a controlled experiment both with and without prediction algorithms (no auto-correct nor suggestions). Our motivation is twofold: first, we want our results to be general and not impacted by the prediction system's accuracy. Second, given the recent debate about intelligent text entry systems~\cite{benefit_suggestions, antti_typing_in_the_wild, effects_language_modeling}, we wish to evaluate the benefits of autocomplete and suggestions separately. Thus, \name performance with OOV words is evaluated separately.

\subsection{Baseline}
As a baseline, we consider three text entry techniques adapted for eyes-free text entry on touchpads: tap typing (BlindType~\cite{blindtype}), gesture typing (i'sFree~\cite{isfree}) and cursor-based typing~\cite{remotes2, blindtype}.
However, to the best of our knowledge, the cursor-based method is the only technique comparable to \name in that it allows entering out-of-vocabulary words: gesture typing (i'sFree) only supports the entry of words within its lexicon~\cite{gesture_typing, isfree, indirect_gesture_typing} and, regarding tap typing, users cannot reach the precision required when eyes-free~\cite{blindtype} and techniques such as BlindType rely on a lexicon forbidding OOV words to compensate users's lack of precision.
Therefore, we focus our analysis on the cursor-based method using the results reported by Lu et al.~\cite{blindtype}, and report the results of i'sFree~\cite{isfree} and BlindType~\cite{blindtype} for reference.
To allow a comparison of our results, we strove to follow the experimental protocol proposed by Lu et al.~\cite{blindtype} in their second user study (§6) and Zhu et al.~\cite{isfree} in their second experiment (§5), as closely as possible. We used the same dataset, the same instructions, and the same number of sentences. The differences that remained had to do with differences in hardware and how the technique works, and are all reported in parenthesis and highlighted in italics below.

\subsection{Participants and Apparatus}
We recruited \textit{12} participants (16~\cite{blindtype}, 18~\cite{isfree}) different from Study 1 (21 to 33 age range, mean = 26.58, 3 identified as female and 9 identified as male, 3 left handed). Participants rated their familiarity with the QWERTY layout 4 out of 5 on average (\std{1.15}).
Participants sat in front of a \textit{27-inch} display  (50-inch~\cite{blindtype}, 46-inch~\cite{isfree}) and used a \textit{5.9-inch Huawei Mate 10 phone} (4.3-inch~\cite{blindtype}, 5.2-inch~\cite{isfree}) with no display feedback on the smartphone. The experimental software was implemented in Java and communication between the smartphone and a computer connected to an external display was done over UDP using the TUIO protocol~\footnote{\url{https://www.tuio.org/}}.

\subsection{Procedure}
Participants sat in front of a display showing \name. They were asked to hold the smartphone \textit{horizontally} (vertically~\cite{blindtype, isfree}) and to use their \textit{two} (one~\cite{blindtype}, no instruction~\cite{isfree}) thumbs to perform strokes. They had to \textit{put the smartphone under the desk} to ensure that they could not see their hands nor the input device (instructed not to look~\cite{blindtype}, no restrictions~\cite{isfree}). The experimenter then explained the technique and participants could \textit{try the technique} (train by completing 4 sentences in~\cite{isfree}) to make sure they understood how it worked. Participants were asked to transcribe sentences from the MacKenzie and Soukoreff phrase set~\cite{10.1145/765891.765971} as ``quickly and accurately as possible''. Typing was unconstrained; participants could go to the next sentence at any time by pressing `Enter'.

All participants transcribed the same sentences, with their order shuffled to ensure that each sentence appeared only once during the entire session. Each participant completed \textit{7} blocks (5 from~\cite{blindtype} and 4 from~\cite{isfree}), with each block consisting of 8 sentences. In the last 2 blocks, auto-correct and word completion (top-2 suggestions) were introduced, allowing participants to explore these new functionalities before commencing the 6th block. This setup resulted in a total of 672 transcribed sentences (56 per participant).

\subsection{Results and Discussion}
\begin{figure}[t]
    \centering
    \includegraphics[width=0.49\linewidth]{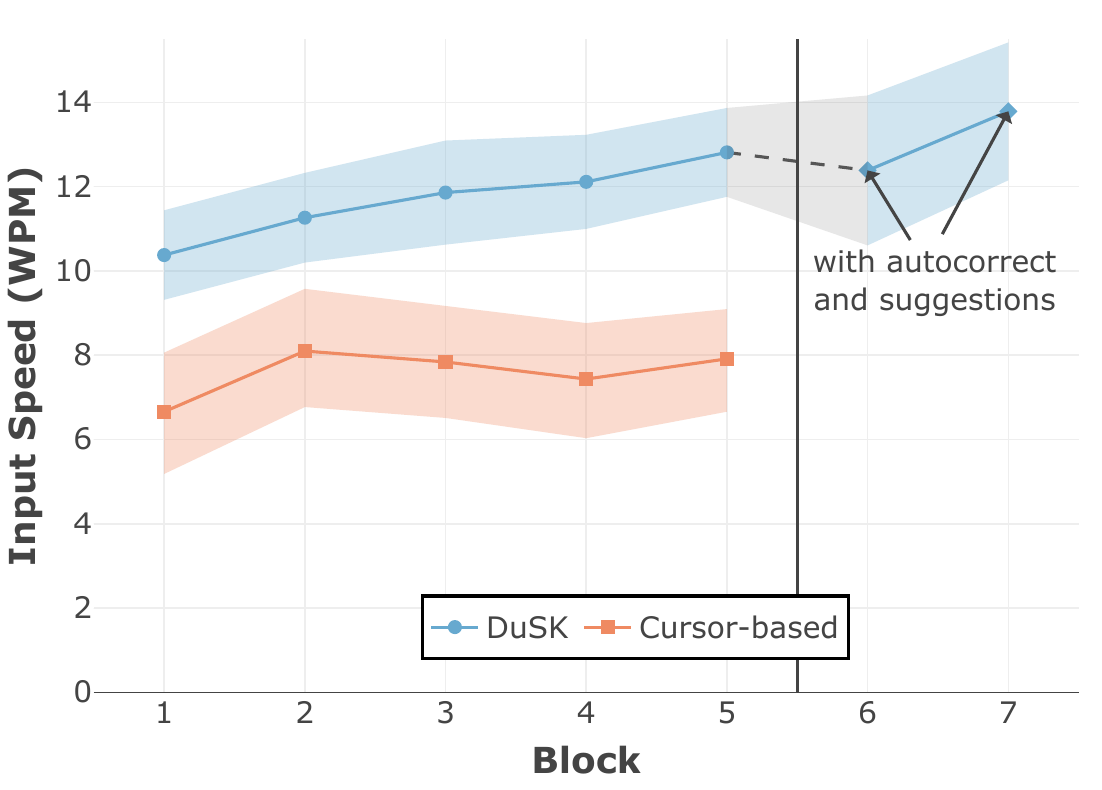}
    \captionof{figure}{Mean and standard deviation of Words-Per-Minutes (WPM) by block for each technique.}
    \label{fig:meanWMP}
\end{figure}

Participants completed the study in 41 minutes on average (\std{9.8}).
We measure words-per-minute (WPM), uncorrected error rate and corrected error rate using Soukoreff and MacKenzie's  equations~\cite{corrected_uncorrected_error_rate_metrics, mackenzie_wpm}.
We used a RM-ANOVA with Greenhouse-Geisser correction when Sphericity was violated and did pair-wise post-hoc comparison using t-tests with Bonferroni correction. For error rate, because the normality assumption was violated, we used a non-parametric Friedman test.
We first present the results of the first 5 blocks, and then report the results of the last two blocks which added autocorrect and word completion.

\textbf{Speed.} Using \name, participants started with an entry rate of 10.38~WPM and reached 12.8~WPM after 5 blocks (ANOVA: significant effect of {\sc Block}, \anovas{6}{66}{14.05},  Block 1 vs Block 5: {\small$p<0.0001$}). Figure~\ref{fig:meanWMP} shows the average WPM for each block.
% TODO: Test effect of block
Interestingly, by the 5th block, participants were still improving and a plateau was yet to be reached.
The 2 participants the least familiar with the QWERTY keyboard (respectively rated themselves 3 and 1 out 5) obtained the two lowest performances (respectively 10~WPM and 10.3~WPM) and the fastest participant had an average speed over the first five blocks of 15.1~WPM.

In comparison, \name is faster than the cursor-based technique on touchpad which was measured to start at a speed of 6.6~WPM and to reach 8~WPM after 5 blocks (two-sample t-test: {\small $p<.001$} for 1st and 5th block). In fact, \name's typing speed on block 1 (10.2~WPM) is superior to the best entry rate achieved with the cursor-based method (8~WPM), suggesting that \name's design is familiar and requires little training to achieve competitive performances.

\textbf{Accuracy.} Consistent with other text entry technique evaluations~\cite{wobbrock_error_metrics}, participants corrected almost all mistakes and left only 1.2\% (\std{5.9}) of uncorrected errors. The corrected error rate was 6.5\% (\std{6.5}). 
A Friedman test did not find a significant effect of {\sc Block} on corrected and uncorrected error rate (\friedman{4}{6.87}{.14} {\small and} \friedman{4}{3.81}{.43}).
The letters `D' and `K' represented 35\% of the letters that participants corrected using backspace. We hypothesize that this is due to participants wanting to type 'Space' and 'Backspace" but incorrectly tapping too high on the touchpad.

\textbf{Benefits of using two thumbs.} Previous research on two-thumb typing showed faster text entry rate when alternating thumb~\cite{two-thumb-entry-model, two-thumb-oulasvirta}. We verify that \name benefits from leveraging two thumbs by measuring the reaction time (i.e. the time between the end of a stroke and the beginning of a new stroke) when switching hands (e.g. the reaction time between stroke A and stroke B, where stroke A was done with the left thumb, and stroke B with the right thumb) compared to using the same hand. We found that participants had significantly faster reaction times when alternating hands as opposed to using the same hand ({\small \mean{439ms} vs \mean{453ms}, p=.017}), suggesting that \name benefits from being two-handed. 

\textbf{Autocorrect and word completion.} In block 6 and 7, the technique was augmented with autocorrect and word completion (top-2). Autocorrect is especially interesting as it allows participants to be less precise and potentially perform strokes faster. However, results showed that autocorrect was used for only 1.7\% of the words. Word completion had more success and participants used it to finish entering 70\% of the words.
%TODO: t-test if average stroke time was significantly different
%TODO: Look at antti's paper, is it lower than normal typing?
We hypothesize that this low adoption of autocorrection is due to the order of the blocks. Participants, being used  to correcting errors right away, were less likely to use the predictive features in the last blocks.

Interestingly, the last block resulted in an increase in speed to 13.8~WPM and a decrease of the uncorrected error rate to 0.59\% but these differences were not significant when compared against the last block without predictions (respectively, {\small$p=.445$} and {\small$p=1$}). Moreover, it is unclear if these performances were due to more training, or because of word completion and autocorrect. 
The benefits of predictive features are investigated in more depth in section~\ref{deconstruct_perf}.

\textbf{Participants comments.} Overall, participants were positive about the technique. 
4 participants commented about the layout, mentioning that they would prefer a different arrangement of space, backspace and enter keys, and that they solicited the left hand more than the right hand (which is essentially a known flaw of the QWERTY layout~\cite{how_we_type}). 2 participants mentioned that the transfer function was too slow, and that they could have performed better with a faster one.
Finally, 2 participants commented that predictions during the last two blocks made typing faster while one participant said he could not split his attention and therefore could not look at the suggestions and make use of them. This is on par with recent findings that debate the benefits of suggestions~\cite{antti_typing_in_the_wild, benefit_suggestions}
\section{Study 3 - \name for OOV words} \label{study_oov}
The phrase set from MacKenzie and Soukoreff~\cite{10.1145/765891.765971} used during Study 2 is commonly used to evaluate and compare text entry techniques, but contains few OOV words~\cite{velocitwatch}, making it non-ideal to observe possible effects of OOV words on the performance of our technique. Since \name relies on the same input dynamic to enter in-vocabulary and OOV words, we expect the results presented in Study 2 to also apply to OOV words.
We verify our hypothesis by running another study to test \name on a more difficult phrase set, including a much higher rate of OOV words. This new experiment used the exact same apparatus and procedure as Study 2 except for the participants and the dataset, as described below.

\subsection{Dataset}
In VelociWatch~\cite{velocitwatch}, Vertanen et al. created a phrase set to contain a high rate of OOV words (with at least one OOV word per sentence) while still being memorable for the purpose of a transcription task. We used this phrase set to assess participants' performance with \name when faced with OOV words.
For the rest of this section, we will refer to the phrase set used in this study as the \textit{OOV phrase set}, as opposed to the \textit{IV phrase set} used during Study 2.

\subsection{Design and Participants}
We used a between-subject design with {\sc Phrase Set} (OOV or IV) as the independent variable and input rate and uncorrected error rate as dependent variables.
We recruited 6 participants different from Study 1 and Study 2 (24 to 42 age range, mean = 30, 3 identified as female and 3 as male). Participants rated their familiarity with the QWERTY layout  3.5 out of 5 on average (SD=1.64).

\subsection{Results and Discussion}

\begin{figure}[t]
    \centering
        \includegraphics[width=0.5\linewidth]{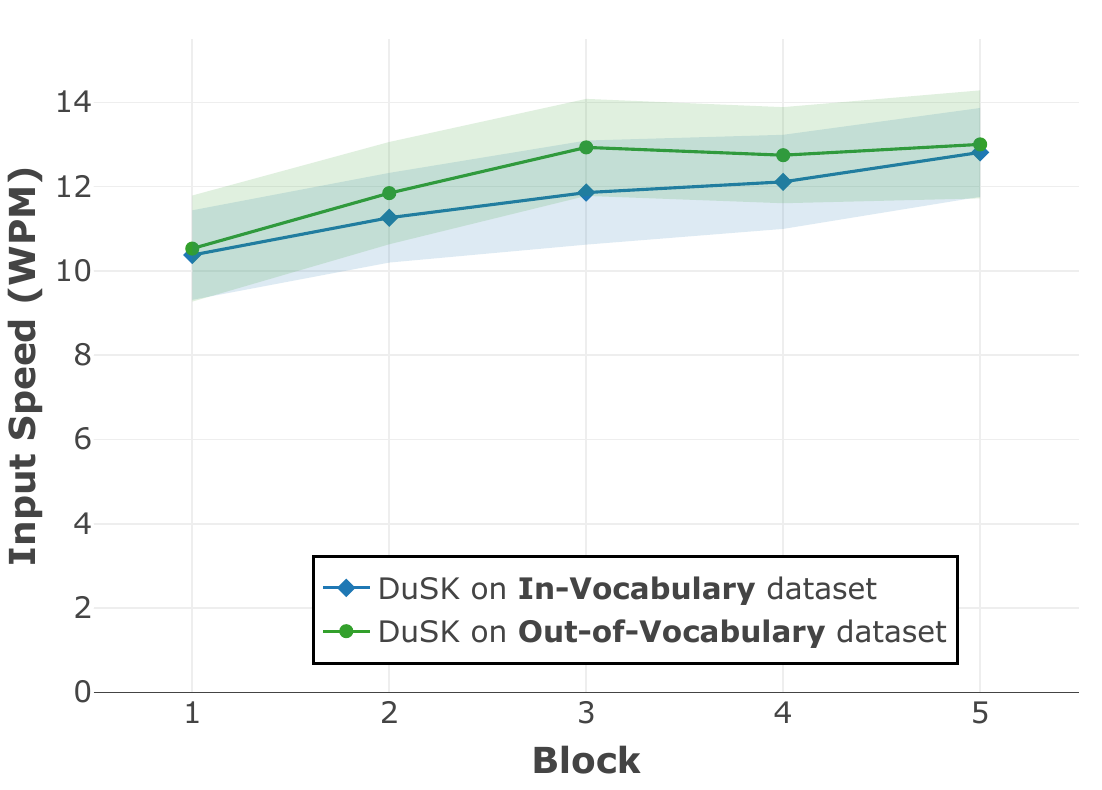}
        \captionof{figure}{Comparison of DuSK's input speed on two different phrase sets: In-Vocabulary~\cite{10.1145/765891.765971} and Out-of-Vocabulary~\cite{velocitwatch}.}
        \label{fig:WPM_oov}
\end{figure}

Figure~\ref{fig:WPM_oov} shows participants' input rate with \name when transcribing sentences from the OOV phrase set and the IV phrase set.
On average, the new set of participants had a similar input rate (\mstd{12.21~WPM}{2.58}) and uncorrected error rate (\mstd{1.76\%}{5.85}) as the participants from study 2, despite entering sentences from a challenging phrase set containing a high rate of OOV words~\cite{velocitwatch}.
% Dataset F(1, 16) =  0.54, p = .471, getasq = .03  
An ANOVA did not reveal a significant effect of the phrase set on input rate (\anova{1}{16}{.54}{.471}) and a Mann-Whitney test did not find a significant effect of the phrase set on uncorrected error rate (\mannwithney{-.14}{.91}).
% Z = -0.14056, p-value = 0.9109

Our result confirms that \name exhibits similar performance on OOV words. We attribute these results to 1) the input dynamic being identical for both OOV and IV words; 2) \name's reliance on unambiguous actions that are accurately performed even without looking at the device. 
% In contrast, other eyes-free text entry techniques either do not provide any solution to enter OOV words~\cite{blindtype, isfree} or result in significantly lower entry rate~\cite{blindtype, remotes2}.

\section{Predicting Expert Performance}
\label{deconstruct_perf}
The previous sections showed that \name outperforms the cursor-based text entry method with and without training. Our results suggest that participants are still improving after 40 minutes. In this section, we use a theoretical model to examine the peak performance that could be reached by an expert using \name. Our model is inspired by the two-thumb text entry model proposed by MacKenzie and Soukoreff~\cite{two-thumb-entry-model}, which has been adapted with success to a wide range of text entry techniques~\cite{two-thumb-clarkson, two-thumb-dunlop, how_we_type_mobile}. The idea is to predict the performance by looking solely at the linguistic and motor components. Below, we describe how we constructed the model and report the predicted peak performance.

\subsection{Model}
Entering a character with \name is done by using strokes or taps. We note $t_{key}(k)$ the time to select a key, $k$, and estimate its value based on the data collected from Study 1. For each character entered using a stroke, we compute the median time it took participants to stroke toward this character and add the time to tap in place (measured to be 127ms on soft-keyboards~\cite{high-perf-kb}). For taps, we compute the median time it took participants to tap the zone containing the key, measured from the time the visual stimuli was displayed to the time the finger up event was received by the phone. Because participants were reacting to a visual stimuli in Study 1, we subtract 230ms (i.e. the approximate human reaction time to a visual stimuli~\cite{reaction_time1, reaction_time2}) to only consider the time to tap.

We then follow the two-thumb typing model proposed by MacKenzie and Soukoreff~\cite{two-thumb-entry-model} with two differences: 1) The left thumb is always used for SPACE given that the key is located on the left; 2) The time to select one key repeatedly with the same finger (referred to as $t_{REPEAT}$ in \cite{two-thumb-entry-model}) depends on the key being selected (e.g. the time to repeatedly select `Q' is going to be longer than for `S' given than the stroking distance is different). Thus, the total time $T_n$ it takes to reach and enter the \textit{n}th letter in a word is:
\[
    T_n = T_{i-1} +
\begin{dcases}
     t_{key}(key_{i}),& \text{if } thumb_{i-1} = thumb_i\\
    \frac{t_{key}(key_{i})}{2},              & \text{otherwise}
\end{dcases}
\]
With $thumb_i$ the thumb used to select the \textit{i}th character,  $thumb_0$ being the left thumb as we assume all words are preceded by a SPACE key~\cite{two-thumb-entry-model}.
%Below, we give an example of using our model to compute the time to enter a word.
%Entering the word 'dusk':
%$t_D + t_U/2 + t_S/2 + t_K/2 + t_SPACE/2$
%D was preceded by a space, therefore same thumb
%U was preceded by D, alternating thumb, we divide by two

%, we assign thumbs to letters by assuming the same correspondence than touch typing. Unlike ~\cite{two-thumb-entry-model}, we do not have uncertainty on the thumb pressing the SPACE key; \name's SPACE key is located on the left side, thus assumed to always be selected using the left thumb.

%Present the model: Use stroke times from study 1 (why?) + strokes are position independent so just use stroke time. Taps are not position independent, so need to use Fitts (but then mixing empirical and theoritical data).

\subsection{Prediction}
Using our model, we calculate that it takes 39,572,285 seconds to enter the 103,183,327 characters of the 17,823,575 words from the \textit{American National Corpus}~\cite{ANC}.
Following the approach proposed by MacKenzie and Soukoreff~\cite{two-thumb-entry-model}, the model predicts a peak expert performance of \name at 31.3~WPM. However, this prediction should be viewed as an approximate upper-bound considering the limitations of the original model it is based on: first, the model is \textit{risk-less} and does not take into account the cost of error correction~\cite{cost_error_correction} or participants typing slower in order to avoid costly errors~\cite{aversion_costly_typing_errors}. Second, the time estimates were obtained empirically from non-experts and might differ in a typing task.

One open question is how good of a prediction of peak performance the model represents. To sanity check the model, two authors trained with \name for two hours prior to entering the first 40 sentences of Study 2. Both authors maintained a text entry rate of more than 23~WPM without auto-correction (respectively 29.9~WPM and 23.6~WPM with an uncorrected error rate of 1\% and 2.6\% respectively). Given this result, the model's prediction of peak performance of \name represents a reasonable estimate.

\vspace{-0.7em}
\section{Discussion}
% Summarise important findings:
% 1. Study 1 trade-off of gestures eyes-free
% 2. Pilot studies lead us to DuSK which is familiar and does not alter input dynamics and therefore allow fast entry of OOV words
% 
Eyes-free text input has received significant recent research interest. Both BlindType \cite{blindtype} and i'sFree \cite{isfree} are contemporary, efficient, eyes-free text input techniques, i.e. are techniques where an invisible keyboard is used to support text entry on distant displays. However, both BlindType and i'sFree are restricted to only in-vocabulary words, and this limits their utility for contexts where OOV words (e.g. web pages, passwords, some proper names) may need to be inputted. The only option to input OOV via BlindType and/or i'sFree is to display a keyboard and switch to character-by-character, accurate targeting text entry.
If the context of use for an eyes-free text entry technique is a smartphone as text entry device and the target, distant display is a physical display in the world such as a smart tv, then it is possible for a keyboard to be displayed on the smartphone and users can look back and forth between displays. However, if the text entry device is, for example, a touchpad equipped remote or if the target display is a head-mounted display, then some other mechanism for text input must be adopted.  For example, the user may need to stop text input, invoke a virtual keyboard, and then type character-by-character.

To address the inability of contemporary eyes-free text entry techniques to support OOV input, this paper explores an alternative text entry mechanism that can support deterministic input by using location-independent, directional gestures. Study 1 found that participants are accurate when performing location-independent gestures despite not seeing their hands nor the touchpad. Our participants also had different preferences and accuracy based on the thumb used, confirming and generalizing the results of Kin et al.~\cite{10.1145/1993060.1993066} to eyes-free contexts. This large set of directional thumb gestures can inform the design of a broad range of eyes-free interactions. 

We leveraged the results of Study 1 to design our eyes-free, directional text entry technique.  In \name, we favored strokes of different length to disambiguate characters that were aligned (e.g. an 'a' and 's' on a keyboard), as informal studies revealed that alternative such as compound strokes for post-hoc disambiguation increased cognitive load on participants. While length disambiguation is unusual for marking-menus (which were originally conceived of as a scale-independent~\cite{marking_menu} invocation techniques), in an eyes-free context in which location-dependent gestures are difficult to perform leveraging size provided a practical way for participants to disambiguate characters.  Leveraging two sizes of strokes effectively doubles the input space along principle axes of input.

Given the above observations, \name can be viewed as a soft keyboard in which the requirement of precisely aiming into the bounding box of keys has been relaxed by relying on strokes and bi-level distances. 
Using \name, a character is accessed through a single unambiguous and location-independent action.
Consequently, \name allows the entry of text, including OOV words, even when the input device and users' hand is hidden (VR and SmartTVs). Additionally, \name was designed to take advantage of both hands; alternating hands reduces reaction times, finger travel distance and also allows for \textit{finger preparation} that happens when the other finger is preparing for the next key in parallel~\cite{how_we_type_mobile, antti_kalq}.

Study 2 showed that novices and experts are faster with \name than with the default and only technique allowing OOV words with SmartTVs: the cursor-based technique.
However, as expected, some lexicon-based methods~\cite{isfree, blindtype} provide faster performance for novice users when typing is restricted to in-dictionary words.
Considering these differences, we believe that novice users could benefit from an hybrid solution combining the strengths of both methods; novice users could use predictive methods to take advantage of their performance and switch to \name whenever they need to enter passwords for example. Switching from one technique to the other can be as simple as rotating the device \ang{90}: portrait mode to use a predictive method such as i'sFree~\cite{isfree} (which was designed for this mode) and landscape mode to use \name. This solution has the advantage of slowly building users expertise with \name.  

One open question with any novel text input technique is what the maximum performance supported might be. Leveraging an established predictive model of text entry \cite{two-thumb-entry-model}, we calculate an approximate peak expert input speed of 31.3 WPM.  As we note based on an informal evaluation by two expert users, this peak text entry rate seems a reasonable bound on performance.  This theoretical peak performance also compares well with text entry rates observed for word gesture keyboards and soft keyboards in the wild \cite{antti_typing_in_the_wild}.  As such, as users develop expertise with \name, and depending on the frequency of OOV words as input, \name presents a useful alternative text entry mechanism, particularly given the constraint of current, high-speed, eyes-free text entry techniques to lexicon input \cite{blindtype,isfree}.

\vspace{-0.6em}
\subsection{Limitations and Future work}
\textit{Touchpad size.} We evaluated \name using a smartphone with a common touchscreen size (5.9 inches). Given the variability of touchpad sizes on different controllers, an interesting direction for future work would be to measure the impact of smaller touchpads on the performance of \name. Granted that the touchpad can be held with two hands, adapting \name would be a matter of tweaking the transfer function. 

\textit{Special characters.} A fully-featured keyboard should support entering special characters, numbers and uppercase letters, which we did not investigate. Because \name strives to leverage users knowledge of the QWERTY layout, strokes are assigned to characters based on the arrangement of the keys, resulting in a large number of strokes left unexploited (from \ang{180} to \ang{225} with the left thumb, and \ang{270} to \ang{360} with the right thumb, see Figure~\ref{fig:mean-angle}) that could fit at least 6 more items. Future work could investigate the use of these strokes to accommodate for direct access to common special characters such as commas and periods. If more characters are needed, a mode-switching key that turns letters into special characters could be added (similar to how uppercase and special characters are accessed on smartphones' soft keyboards).

\textit{Other input devices.} Essentially, \name needs 4 degrees of freedom (DoF) for strokes (2 per thumb) and 3 buttons to select 'Space', 'Backspace', and 'Enter' (although these buttons could arguably be associated to strokes). We artificially augmented the number of DoF of the touchpad by dividing its contact area. Numerous input devices with similar capabilities could support \name and might benefit from its use. Future work could investigate the performance of \name with such devices (HTC Vive controllers' touchpads, dual-joysticks game controllers, etc.).

%Open the discussion on rather Two-handed vs single handed.
%No need for peripheral vision => not the case for i'sfree.
%No visual feedback =>

%\textit{Eyes-free.} Previous work used the term eyes-free to designate techniques to enter text without looking at the input device rather than with no visual feedback: i'sFree~\cite{isfree} shows the inputted text and the top suggestions and the keyboard is shown if necessary. Similarly, BlindType~\cite{blindtype} also shows the text entered, suggestions and the keyboard at all time. By this definition, \name could be qualified as eyes-free, although, because of the ambiguity, we deliberately did not use this term throughout this work. \name was evaluated with visual guidance (i.e. a keyboard was showing on the distant display, and participants strokes were drawn on top of it) but participants could not see their hands, nor the input device. Although \name could be tested without any visual guidance, the only benefit would be to save space on-screen, considering that recent work on marking-menus suggests that it does not help promoting novice to expert transition~\cite{jay_marking_menu_delay}, in contradiction with previous beliefs~\cite{marking_menu}.

%\input{tex/futurework}
\section{Conclusion}
We present \name, a technique to support expressive eyes-free text input on touchpads. We first compared users' thumb-based strokes and taps precision and speed when unable to see the touchpad. \name was then designed for eyes-free and bi-manual interaction through an iterative process. We found through an experiment that our proposed design outperforms current deterministic solutions and that expert users can reach performances approaching that of sighted tap typing on smartphones. We believe that, in contexts where the input device is decoupled from the output device such as SmartTVs and Virtual Reality, the ability to enter out-of-vocabulary words is critical; \name provides a solid alternative to lexicon-based techniques and can serve as a replacement to the widespread cursor-based method.
\section*{Acknowledgment}
We thank our late supervisor, Dr. Edward Lank, for his invaluable guidance and support on this work. His contributions were essential to this research, and He is deeply missed.

\bibliographystyle{ACM-Reference-Format}
\bibliography{references}

\end{document}